\documentclass{aastex63}

\newcommand{\kms}[0]{km~s$^{-1}$~}
\newcommand{\msun}[0]{M$_{\Sun}$}
\newcommand{\msyr}[0]{M$_{\Sun}$~yr$^{-1}$~}

\defcitealias{Tanner17}{Tanner17}
\defcitealias{CC85}{CC85}
\defcitealias{JA71}{JA71}

\received{\today}
\revised{}
\accepted{July 3, 2020}

\submitjournal{ApJ}

\shorttitle{Impeding Galactic Outflows}
\shortauthors{Tanner}

\graphicspath{{./}}

\begin{document}

\title{Massive Galaxies Impede Massive Outflows}

\correspondingauthor{Ryan Tanner}
\email{ryan.tanner@nasa.gov}
\author[0000-0002-1359-1626]{Ryan Tanner}
\affiliation{NASA Goddard Space Flight Center\\
Greenbelt, Maryland, 20771, USA}
\affiliation{Augusta University\\
Augusta, Georgia, 30912, USA}

\begin{abstract}

%The kinematics and composition of a galactic outflow can be greatly affected by the stellar mass of the host galaxy. 
A set of 66 3D hydrodynamical simulations explores how galactic stellar mass affects three-phase, starburst-driven outflows. Simulated velocities are compared to two basic analytic models: with (Johnson \& Axford 1971) and without (Chevalier \& Clegg 1985) a gravitational potential. For stellar mass $<10^{10}$ \msun, simulated velocities match those of both analytical models and are unaffected by the potential; above they reduce significantly as expected from the analytic model with gravity. Gravity also affects total outflow mass and each of the three phases differently. Outflow mass in the hot, warm, and cold phases each scale with stellar mass as $\log M_*=$ -0.25, -0.97, and -1.70, respectively. Thus, the commonly used Chevalier \& Clegg analytic model should be modified to include gravity when applied to higher mass galaxies. In particular, using M82 as the canonical galaxy to interpret hydrodynamical simulations of starburst-driven outflows from higher mass galaxies will underestimate the retarding effect of gravity. Using the analytic model of Johnson \& Axford  with realistic thermalization efficiency and mass loading I find that only galaxy masses $\la10^{11.5}$ \msun~can outflow.

\end{abstract}


\section{Introduction}\label{sec:intro}

Much observational and modeling work has established scaling relations between host galaxy properties and the kinematics and content of galactic ``superwind" outflows  \citep{2005ARA&A..43..769V,2017hsn..book.2431H,2018Galax...6..138R,2018ApJ...854..110Z,2020arXiv200207765V}.
Influential host galaxy properties include star formation rate (SFR), SFR density, UV, optical and infrared emission and absorption, galaxy mass, thermalization efficiency, and mass loading.

A still popular galactic outflow model was proposed by \citet[][hereafter \citetalias{CC85}]{CC85} with wind generated by thermal overpressure of a nuclear starburst
\citep{2005ARA&A..43..769V,2017hsn..book.2431H}. 
CC85 argued that gravity barely affects the outflow. 
Subsequent wind models added radiation pressure \citep{2009MNRAS.396L..90N,2015MNRAS.449..147T} or 
cosmic rays \citep{1975ApJ...196..107I,2020MNRAS.492.3465H} or all three \citep{2020MNRAS.492.3179Y}, to  improve upon CC85. 
The less employed \citet[][hereafter \citetalias{JA71}]{JA71} model
also uses thermal over-pressure to generate a wind but adds a spherical gravitational potential. 

Variants of the CC85 model have determined the inputs and subgrid parameters of galactic scale hydrodynamical simulations \citep{StricklandStevens,CooperI,Tanner17,2020MNRAS.492.3179Y,2020arXiv200210468S}. 
\cite{2020MNRAS.492.3179Y} review how the models affect the mass loading rates, mass and energy outflow rates, and terminal wind velocities by following several gas phases \citep{1999ApJ...523..575L,2009ApJ...700L.149V,2013ApJ...774..126M}. 
But as \cite{2020arXiv200210468S} show, different CC85 model variants can influence differently each phase of a galactic outflow.

Different motions of the each phase arising from differing determining factors require that we treat each separately \citep{2016ApJ...822....9H,2019ApJ...878...84M,2020arXiv200210468S}. 
In \citet[][hereafter \citetalias{Tanner17}]{Tanner17} I discussed my 3D simulations that reproduced and explained some of the observed scaling relations between galactic wind velocities and the SFR.
In this paper I use my simulations to compare the predictions of the basic CC85 model to that of the JA71 model over a range of galaxy mass. 
Section \ref{sec:setup} explains my setup, and carefully defines the term mass loading because it is used differently by different authors. 
I then explain how I measure the velocities of the three phase outflow, and in
\S\ref{sec:velocity} show how they are better explained by the JA71 model. 
In \S\ref{sec:massout} I show how my simulations reproduce the observed negative correlation between mass outflow rate and galaxy mass. 
In \S\ref{sec:maxmass} I explore the consequences of the JA71 model \textit{vs} the CC85 model on the predicted maximum galaxy mass at which galactic outflows can form.

\section{Simulation Setup}\label{sec:setup}

To explore how outflow velocities scale with galaxy mass I setup a series of hydro simulations using the Athena \citep{Stone-Athena} MHD code with magnetic fields turned off. 
The canonical model is an M82-mass galaxy \citep{Tanner17,CooperI}. 
I model its disk by adding a Plummer-Kuzmin potential \citep{1975PASJ...27..533M} 
\begin{equation}\label{phidisk}
\Phi_{\rm disk}(r,z)=-\frac{GM_{\rm disk}}{\sqrt{r^{2}+(a+\sqrt{z^{2}+b^{2}})^{2}}}
\end{equation}
to a spherical King model
\begin{equation}\label{phiss}
\Phi_{\rm ss}(R)=-\frac{GM_{\rm ss}}{r_{0}}\left[\frac{\ln\left[(R/r_{0})+\sqrt{1+(R/r_{0})^{2}}\right]}{(R/r_{0})}\right].
\end{equation}
Galaxy parameters for the potentials are the same as in \citetalias{Tanner17}, except that I vary the stellar $M_{disk}$ from $10^{9.6}$ $M_{\sun}$ to $10^{10.6}$ $M_{\sun}$ in steps of 0.2 dex. 
I did not include a dark matter halo; when I included one in my analytic models it did not change any of the relations or conclusions presented in this paper. 
But for completeness a dark matter halo should be included in all future simulations. 
The thickness of the gaseous disk is set by a tanh profile of scale height 110 pc to correct the nonphysical flaring of the disk at large radii as seen in \citet{CooperI} and \citetalias{Tanner17}.
For each galaxy mass I run 11 simulations with CC85 model velocities ranging from 400 \kms to 2400 \kms in steps of 200 \kms, 66 simulations in all.
%\footnote{Only 64 simulations are included in the analysis since the data for simulations with galaxy mass $=10^{10.2}$ \msun, $v_A = 600$ \kms; and $=10^{10.0}$ \msun, $v_A = 1600$ \kms, along with all simulations with galaxy mass $>10^{10.6}$ \msun~were lost when lightning struck the building while the author was transferring files to make a new backup after the first backup hard drive had failed.}. 
I choose the CC85 model velocity and set the thermalization efficiency ($\epsilon$) to 1.0, then calculate mass loading ($\beta$) using Equation \ref{eq:va}.
The initial gas density distribution is set by a semi-random fractal distribution as in \citetalias{Tanner17}. 
Within the starburst region, mass and energy are injected proportional to the initial density, with total energy and mass injection rates set by a Starburst99 model with an SFR of 50 \msyr. 

\subsection{Thermalization Efficiency and Mass Loading}\label{sec:setup:etabeta}

Thermalization efficiency $(\epsilon)$, the fraction 0 to 1 of starburst power output absorbed by the surrounding medium, varies with environment and perhaps time \citep{2003ApJ...594..888F,2005ARA&A..43..769V,2009ApJ...697.2030S,2009ApJ...700..931S,2015ApJ...802...99K}. 
\citet{2003ApJ...594..888F} estimated that immediately after star formation $\epsilon\approx0.1$ but would thereafter fall quickly to $\approx 0.01$.
\cite{2009ApJ...700..931S} measured $\epsilon<0.1$ for star clusters in M82, but \cite{2009ApJ...697.2030S} modeled a range of 0.1-1.0 finding $>0.3$ most likely in M82. 
\citet{2015ApJ...802...99K} found that $\epsilon$ could shift rapidly between 1 and 0.1-0.3 depending on environmental and starburst properties. 
Here I set $\epsilon=1.0$.

The term mass loading is easily confused because it is used in different papers to refer to related but different things. 
\citetalias{CC85} used it to refer to all gas swept up from a star cluster, including mass from stellar winds, supernova ejecta, gas left over from star formation, and any diffuse ambient ISM. In their formulation, mass loading would be

\begin{equation}\label{eq:CC85Mdot}
\dot{M} = \dot{M}_{SN+SW} + \dot{M}_{cold} + \dot{M}_{ISM}.
\end{equation}

Simulations of nuclear starbursts use a sub-grid model to account for sub-parsec scale gas remaining from star formation \citep{StricklandStevens,2009ApJ...697.2030S,CooperI,Tanner17}. 
Mass injected by stellar winds and supernovae $(\dot{M}_{SN+SW})$ is calculated using Starburst99 models \citep{Leitherer1999}. 
Multiplicative mass loading factor $\beta$ accounts for unresolved molecular clouds. 
So total mass injected per timestep is

\begin{equation}\label{eq:Mdot}
\dot{M} = \dot{M}_{SN+SW} + \dot{M}_{cold} = \beta\dot{M}_{SN+SW}.
\end{equation}
\noindent
Evidently, this is a subset of that in \citetalias{CC85} because it does not include the diffuse ISM gas swept up by the wind. 
In my simulations the diffuse ISM is the initial density. 
Together $\epsilon$ and $\beta$ determine outflow velocity as I explain in \S\ref{sec:setup:va}.

Some authors \citep{2015MNRAS.454.2691M,2015ApJ...809..147H,2020MNRAS.tmp..439R} use mass loading to refer exclusively to mass that leaves the galactic disk.
This is a subset of the mass used in \citetalias{CC85} because only a fraction of the wind is directed out of the disk plane. 
That is, $\dot{M}_{out}$ is only a fraction of $\dot{M}$ from Equation \ref{eq:Mdot}, but $\dot{M}_{out}$ adds mass swept up from the ISM. 
However, $\dot{M} \propto \dot{M}_{out}$ with proportionality that depends on the ambient ISM pressure and density, the extent and duration of the starburst, and the opening angle of the outflow. 
To establish the exact relation between $\dot{M}$ and $\dot{M}_{out}$ would require simulations of starbursts on the sub-parsec scale with parameter studies of ambient ISM pressure and clumpiness, the effect of cosmic rays, and radiation pressure \citep{2015ApJ...802...99K,2020MNRAS.492.3179Y,2020arXiv200210468S}.

To avoid confusion I use $\beta$ to refer to mass loading of the wind as defined in Equation \ref{eq:Mdot}, and 

\begin{equation}\label{eq:eta}
    \eta = \frac{\dot{M}_{out}}{\text{SFR}}
\end{equation}
\noindent
to refer to mass loading from the galaxy, with $\dot{M}_{out}$ the mass outflow rate.
How $\eta$ scales to other galaxy parameters is a common diagnostic of galactic wind properties \citep[see references in][]{2018Galax...6..138R}.

Because $\beta$ depends on $\dot{M}_{cold}$ -- essentially a measure of the mass of the cloud not converted into stars -- one can simply assume that $\beta$ is inversely proportional to the star formation efficiency (SFE)

\begin{equation}\label{eq:sfe}
    \beta \propto \frac{1}{\text{SFE}}
\end{equation}
\noindent
The exact relation would require higher resolution simulations \citep[e.g.\ ][]{2018ApJ...853..173K,2019MNRAS.483.3363H}. 
This would imply that $\eta$ is also inversely proportional to the SFE so both $\eta$ and $\beta$ can be proxies for the SFE. 

\subsection{Outflow Velocity}\label{sec:setup:va}

The simple model of CC85 assumes that thermal pressure on the ISM by a star cluster goes into the kinetic energy of the gas. This results in 

\begin{equation}\label{eq:va}
    v_A = v_0\sqrt{2\frac{\epsilon}{\beta}}
\end{equation}
that I refer to as the CC85 model velocity (for the derivation see  \citetalias{Tanner17}). 
Here I set $v_0 = 1894.0$ \kms; the exact value depends on the energy and mass injection, which I find using Starburst99 \citep{Leitherer1999,Tanner17}. 
Simulations from \citetalias{Tanner17} showed that the velocity of the hot wind does not depend on the SFR. This agress with Equation \ref{eq:va} because the SFR cancels out of the equation \citep{2005ARA&A..43..769V}.

\citetalias{Tanner17} kept galaxy mass constant and varied the SFR. 
Here I keep SFR constant at 50 \msyr but vary galaxy mass. 
CC85 assumed that gravity did not important for wind dynamics. 
But JA71 assumed that it was so included gravitational potential energy. 
Based on their results, I start with 

\begin{equation}\label{eq:vaphi}
    \frac{1}{2}\dot{M}v^2 = \dot{E} - \dot{M}\Delta\Phi(r)
\end{equation}
with $\dot{E} = \epsilon\dot{E}_{SN+SW}$ and $\dot{M} = \beta\dot{M}_{SN+SW}$, where $\dot{E}_{SN+SW}$ and $\dot{M}_{SN+SW}$ are the contributions to the ISM from supernova and stellar winds and are linear functions of the SFR. 
Solving for velocity and simplifying results gives

\begin{equation}\label{eq:vg}
    v_G = \sqrt{2\left(v_0^2\frac{\epsilon}{\beta} - \Delta\Phi(r)\right)},
\end{equation}
which like Equation \ref{eq:va} does not depend on the SFR. 
I set $v_A$ and $\epsilon$ and use Equation \ref{eq:va} to calculate $\beta$ which determines the mass input at each simulation timestep. 
My values of $v_A$ were chosen to span observed outflow velocities \citep{2005ApJ...621..227M,2005ApJS..160..115R,2015ApJ...811..149C,2016MNRAS.457.3133C}. 
Note that some choices can yield nonphysical values of $\beta$, which will be addressed in \S\ref{sec:maxmass}. 
For now I compare measured outflow velocities to those predicted from Equations \ref{eq:va} and \ref{eq:vg}. 

\subsection{Simulation Velocities}\label{sec:setup:vcent}

I measure outflow velocities from synthetic absorption lines generated by the method in \citetalias{Tanner17}.
For simplicity I generate only silicon lines. 
I determine the outflow velocity of the multi-phase gas from the Doppler shift of its line center ($v_{cent}$), which is defined as the half width at half line depth. 
To remove possible contributions from stars or disk gas some authors use $v_{90}$, the velocity on the blueward side of the line where the absorption line reaches 90\% of the adjacent continuum level.
Both methods are used to measure outflow velocities  \citep{2005ApJ...621..227M,2005ApJS..160..115R,2009ApJ...692..187W,2012ApJ...759...26E,2014ApJ...794..130B,2015ApJ...811..149C,2015ApJ...809..147H,2016MNRAS.457.3133C,2016MNRAS.457.1257H,2016A&A...588A..41C,2016ApJ...822....9H,2019ApJ...878...84M}. 
I use $v_{cent}$ because it measures well the mass weighted average velocity of the gas for a specific ionization state.

To trace the velocities of the cold, warm, and hot gas I tracked the Si I, Si III, Si VII, and Si XIII ions. 
As noted in \citetalias{Tanner17}, the measured velocity increases with increasing ionization, with biggest jumps between Si II and Si III, and Si XI and Si XII. 
Between Si III and Si X there is virtually no difference in the measured velocities. 

\begin{figure*}[hb!]
\gridline{\fig{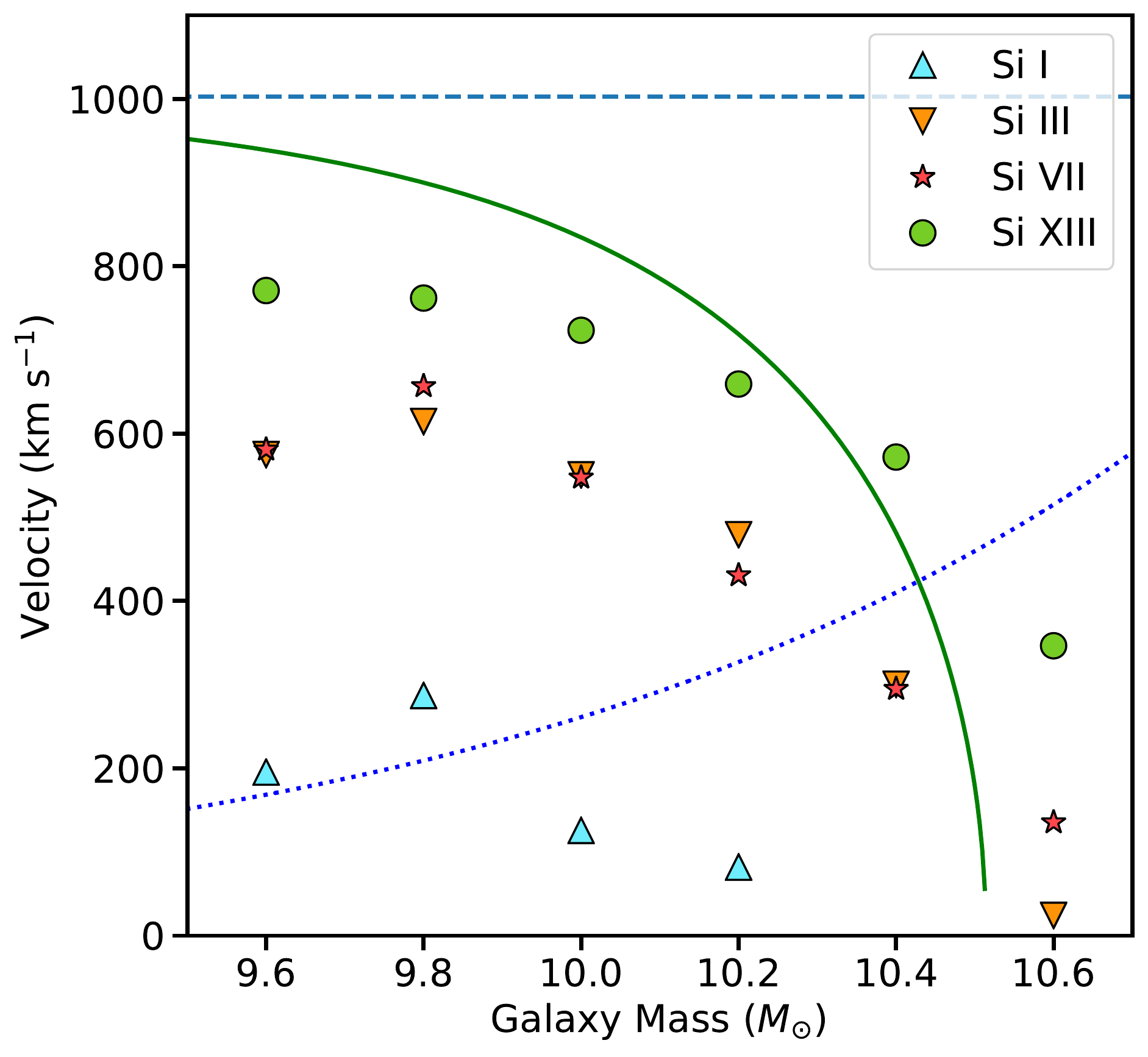}{0.33\textwidth}{}
          \fig{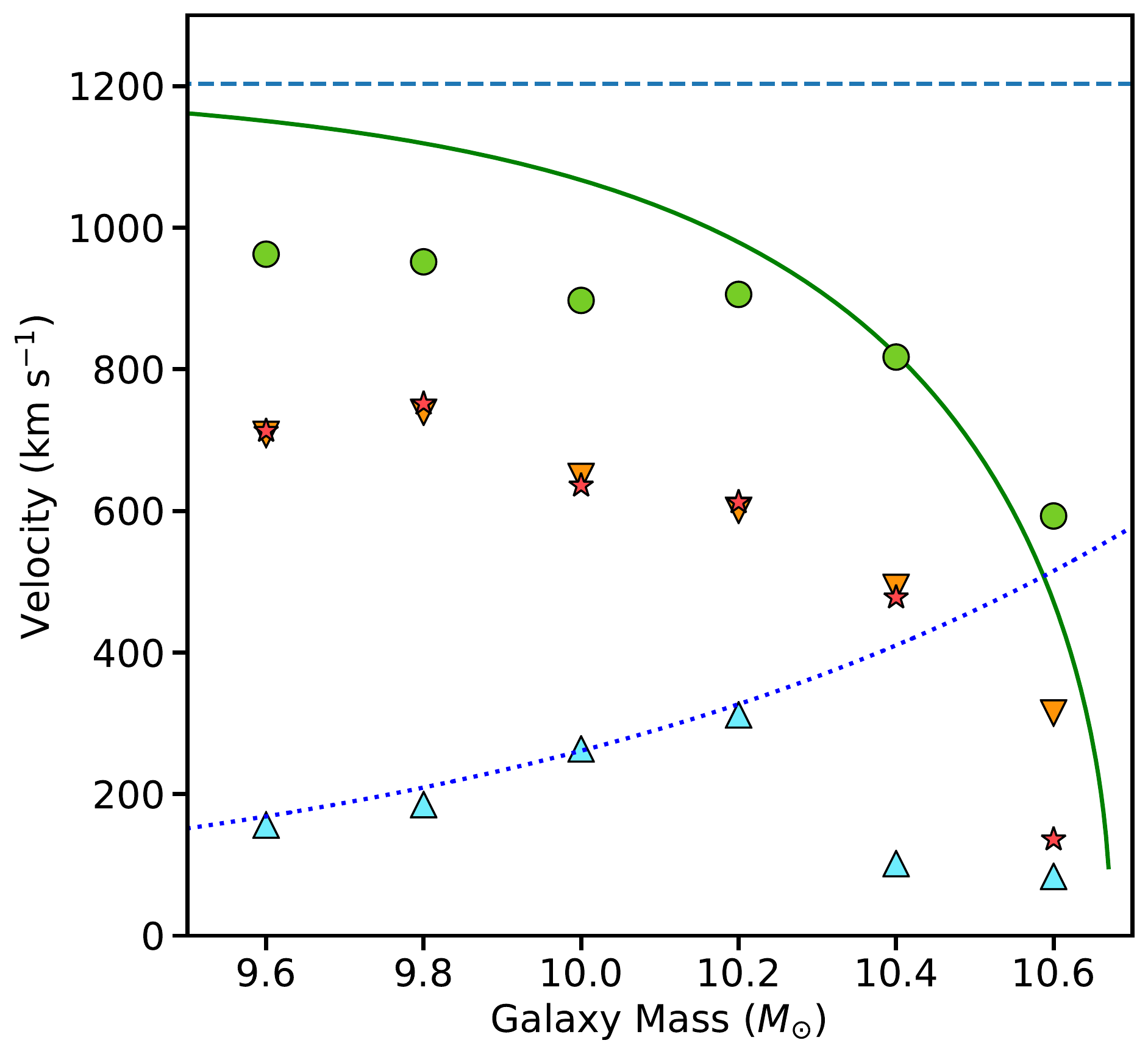}{0.33\textwidth}{}
          \fig{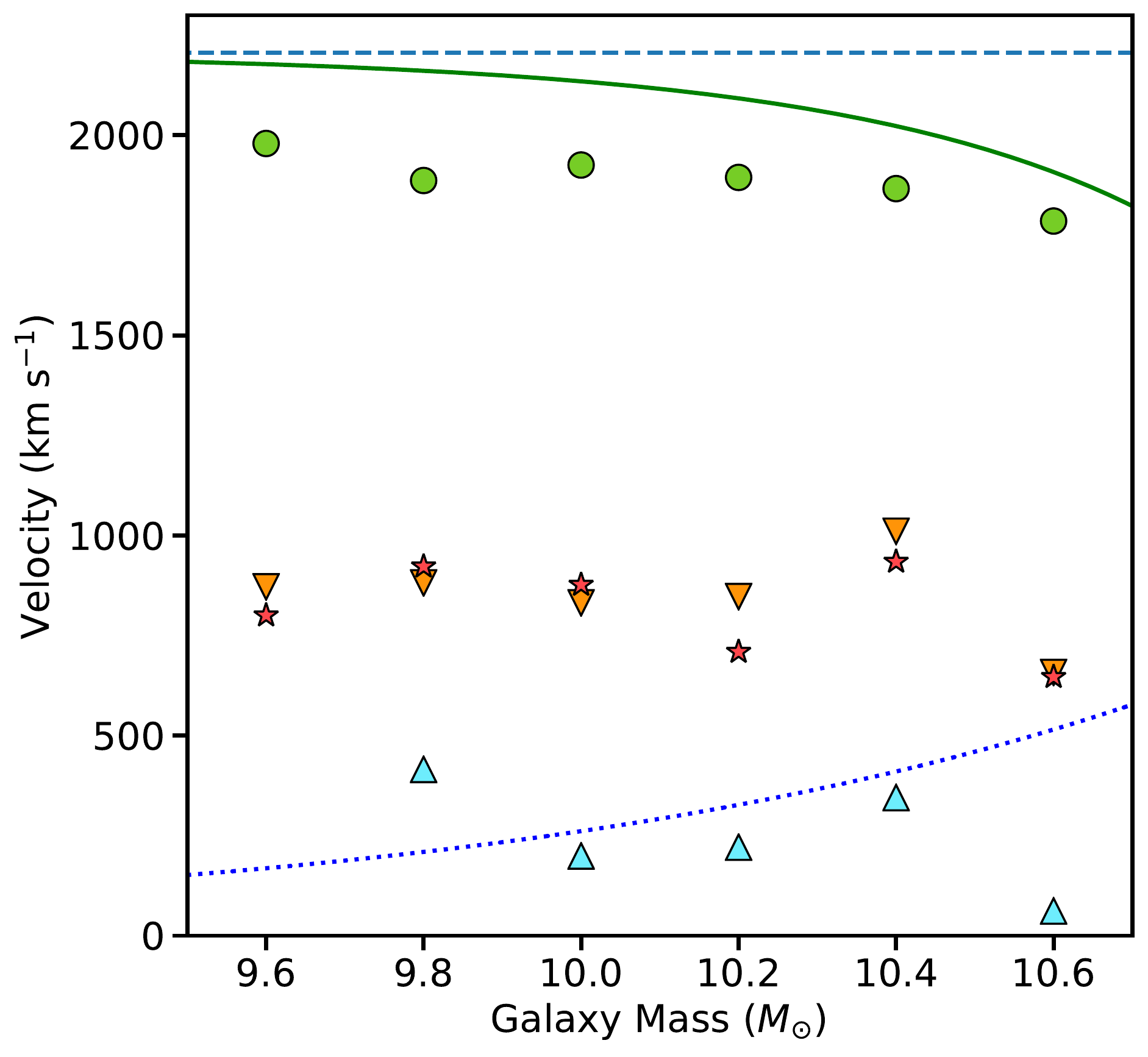}{0.33\textwidth}{}
          }
\caption{Plots of outflow velocity vs. galaxy mass. Outflow velocities are measured by Si I (upward triangles), Si III (downward triangles), Si VII (stars), and Si XIII (circles). CC85 model velocities of 1000 \kms (left), 1200 \kms (center), and 2200 \kms (right) are the horizontal dashed lines, corresponding to $\beta$ of 7.2, 5.0, and 1.5, respectively. Each solid green curve indicates the JA71 model velocity, and dotted blue curve indicates the escape velocity for gas in the galactic nucleus.\label{fig:linev}}
\end{figure*}

\section{Outflow Velocities}\label{sec:velocity}

\citetalias{Tanner17} showed that the velocity of highly ionized gas $(v_{hot})$ was $\sim80-90\%$ of the CC85 model velocity and did not depend on the SFR. 
CC85 model velocities and mine differ because the CC85 model calculates the maximum velocity and I measure the velocity at line center, which provides an average velocity. Measurements of $v_{90}$ are closer to the CC85 velocities, but both $v_{90}$ and $v_{cent}$ show similar trends \citetalias{Tanner17}. My simulations also cool the gas radiatively, but the cooling time for the hot gas is large compared to the simulation time so has small effect.

Here I find that for a single value of $\beta$, $v_{hot}$ decreases as expected with increasing galactic mass. 
While the difference between the original CC85 model and JA71 is small $(< 3\%)$ for galaxies of mass $\leq 10^{9.6}$ \msun, the predicted velocities diverge, sometimes significantly, for increasing galactic mass. 
Figure \ref{fig:linev} shows the results of three sets of CC85 models spanning the galaxy masses tested. 
The CC85 velocities correspond to $\beta$ of 7.2, 5.0, and 1.5 respectively. 
Each plots the outflow velocities predicted by the CC85 model (Equation \ref{eq:va}) and the JA71 model (Equation \ref{eq:vg}). 
The velocities of the three gas phases, $v_{hot}$, $v_{warm}$, and $v_{cold}$, are measured by Si XIII, Si VII or Si III, and Si I lines respectively. 
As noted in \S\ref{sec:setup:vcent} and shown in \citetalias{Tanner17}, there is no significant difference in velocities for ionization states III-X. 
This can be seen in the very similar measured velocities of Si III and Si VII in my simulations. 
Thus either ionization can be a proxy to measure the velocity of the warm phase. 

Simulations with lower $v_A$, corresponding to higher values of $\beta$, show greater divergence between the CC85 and AJ71 predicted velocities. 
Lower mass galaxies can outflow in all three phases.
In higher mass galaxies it is possible for $\beta$ to quench cold and warm phases so that the outflow is primarily hot gas, i.e.\ a galaxy whose SFE is such that the outflow is almost entirely hot X-ray emitting gas with little optical or IR emission. 

Both $v_{hot}$ and $v_{warm}$ follow the trend of the JA71 model, with $v_{warm}$ lower than $v_{hot}$. 
As explained in \citetalias{Tanner17} $v_{warm}$ and $v_{cold}$, unlike the hot gas, depends on the SFR such that  increased SFR increases outflow velocities.
But higher SFR will only increase $v_{warm}$ up to $\sim0.8 v_{hot}$, whereupon it saturates and is flat for increasing SFR. 
The saturation point for $v_{cold}$ is $\sim0.6v_{hot}$. 

In the left and center panels of Figure \ref{fig:linev}, $v_{warm}$ (measured by Si III and Si VII) has saturated so that, as $v_{hot}$ decreases with increasing galaxy mass, $v_{warm}$ decreases proportionally. 
In the right panel the warm gas is below the saturation point and appears to have the same slight downward trend as $v_{hot}$, but the relation is less clear for simulations with similar CC85 model velocities. 
Figure \ref{fig:outvel} in Appendix \ref{app:outvel} plots the velocities from all simulations. 
Below a CC85 model velocity of 1600 \kms, $v_{warm}$ saturates and follows the same trend as $v_{hot}$. 
The measured cold velocities vary much more because the cold gas in the wind clumps \citep{2013MNRAS.430.3235M}, and ram pressure sets the velocity of each clump \citep{Tanner17,2019ApJ...878...84M}.
But in simulations with low $v_A$ and high $\beta$, the cold gas is entirely quenched for high mass galaxies. 
Most $v_{cold}$ measurements are below $v_{escape}$, but because the velocities are measured at the line centers, some cold gas exceeds $v_{escape}$. 

\section{Outflow Mass}\label{sec:massout}

To calculate the mass outflow rates in my simulations I measure the velocity vertical to the disk for each cell relative to the escape velocity ($v_{out} = v_z - v_{escape}$). 
For each simulation I make mass distributions by binning the mass of the cells according to $v_{out}$ using $\Delta = 100$ \kms bins, for all temperatures, and also for mass with the temperature ranges in Table \ref{tab:slopes}. 
This gives the total mass outflow and those in different gas temperature ranges.

\begin{figure*}[h!]
\gridline{\fig{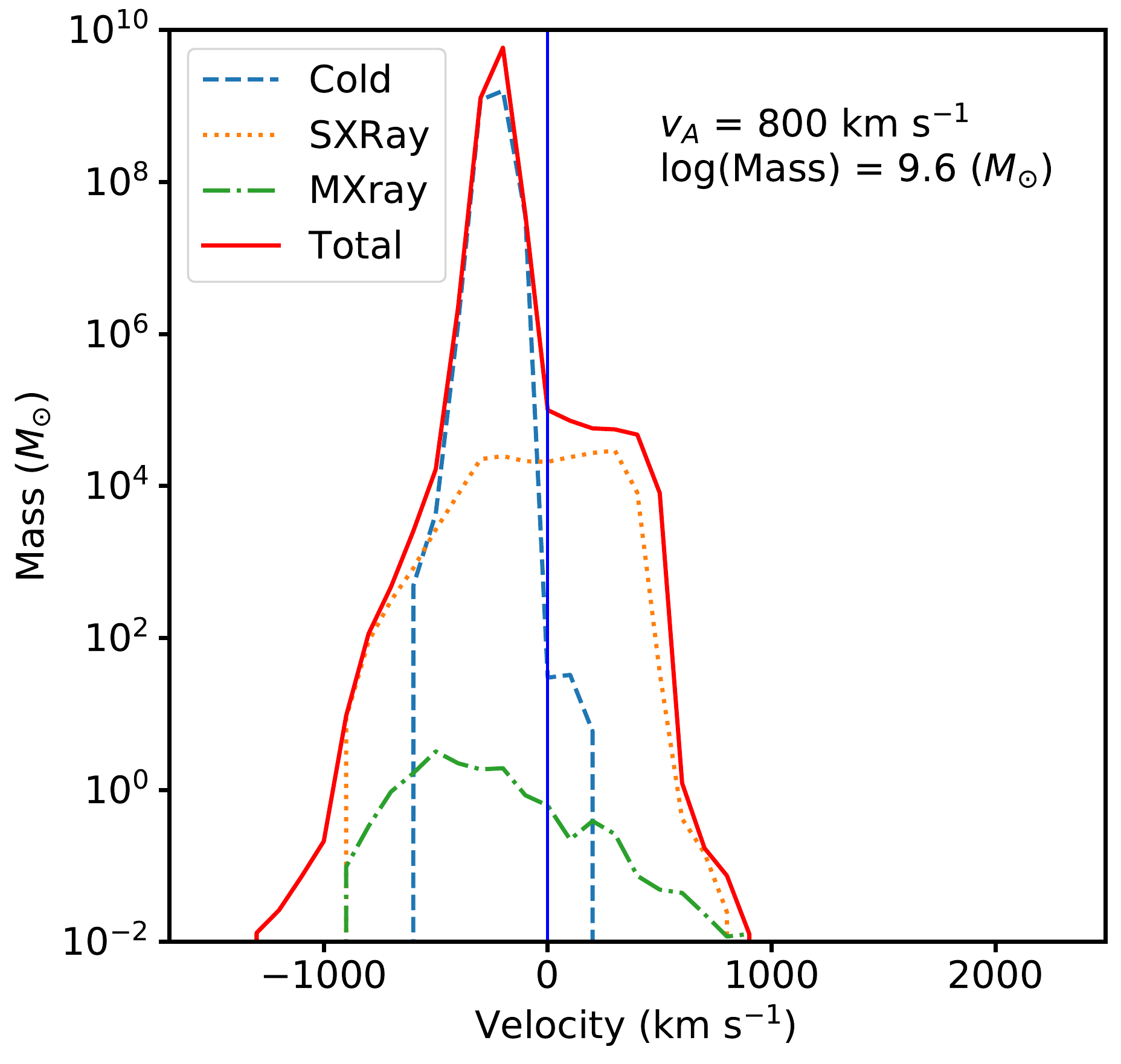}{0.3\textwidth}{}
          \fig{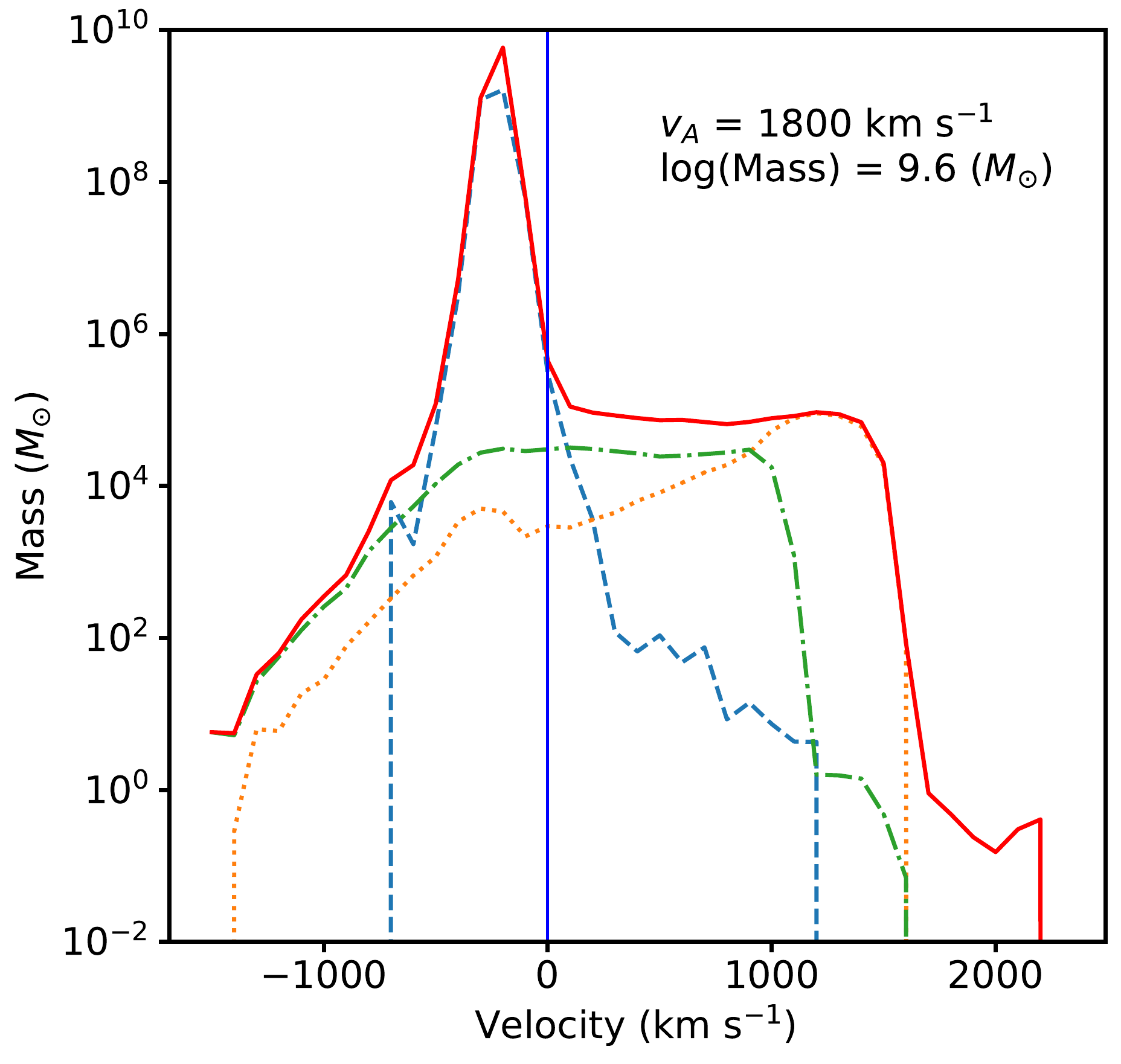}{0.3\textwidth}{}
          \fig{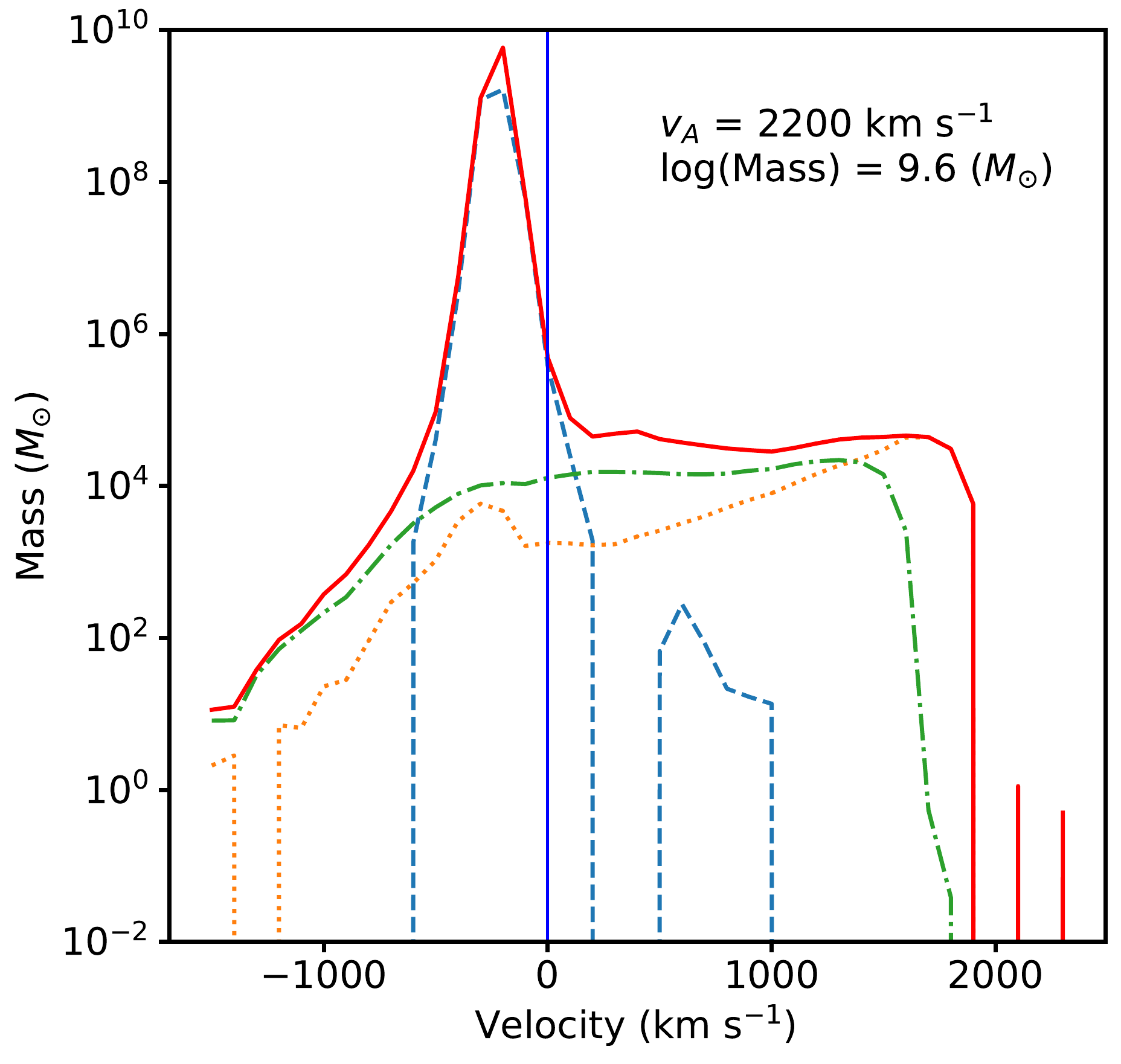}{0.3\textwidth}{}
          }
\gridline{\fig{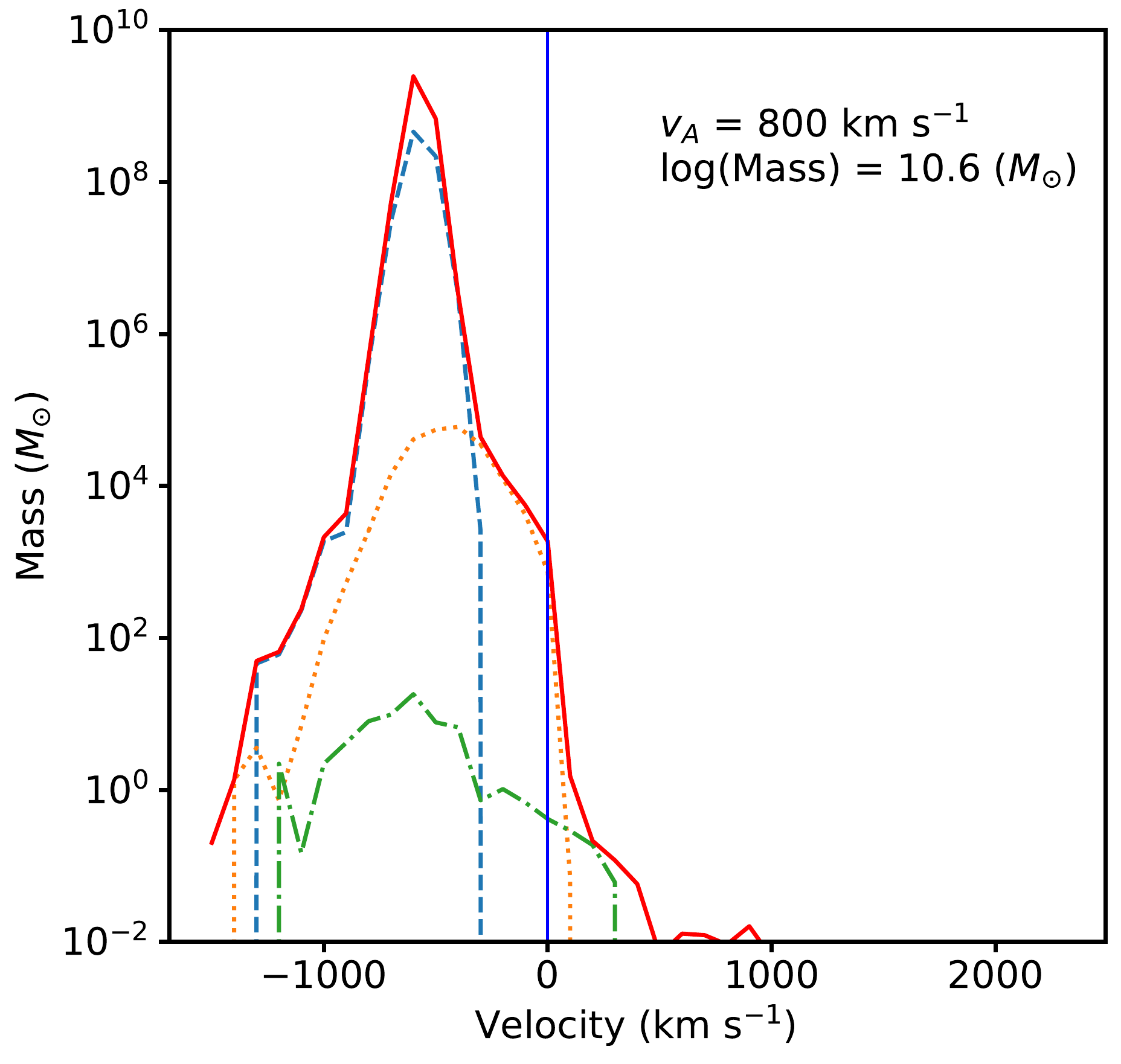}{0.3\textwidth}{}
          \fig{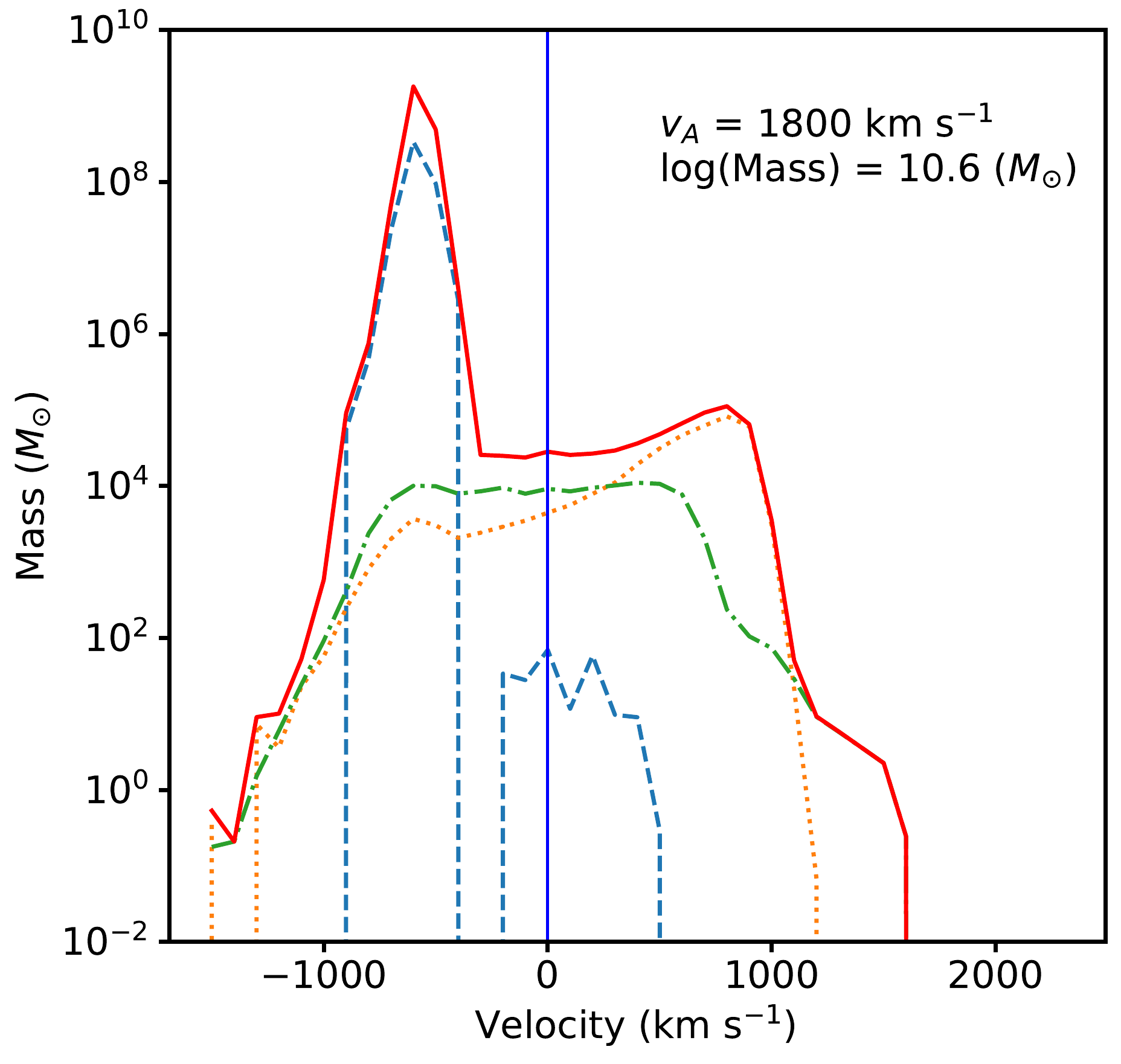}{0.3\textwidth}{}
          \fig{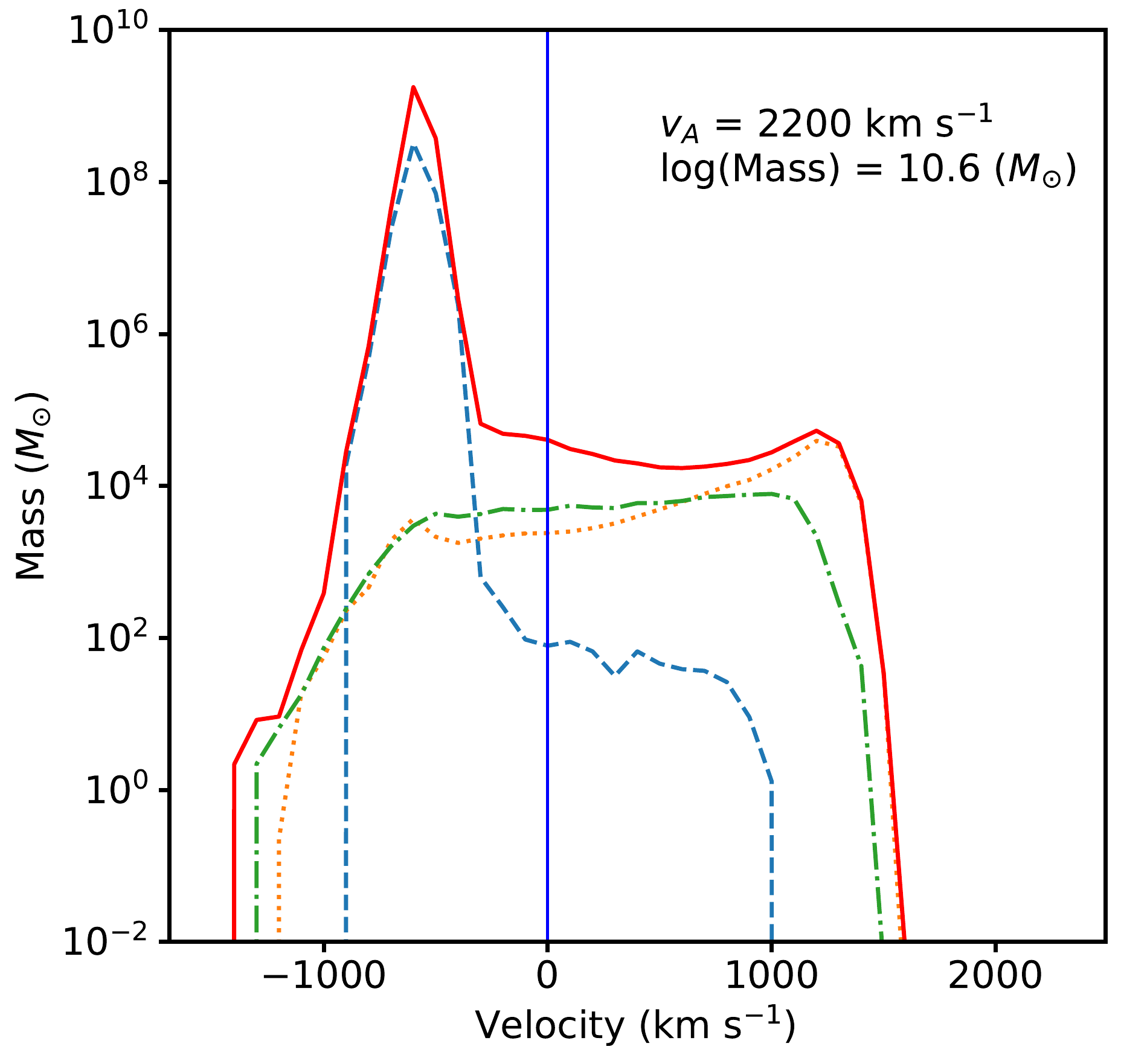}{0.3\textwidth}{}
          }
\caption{Example mass distributions binned according to $v_{out}$ for two galaxy masses (top row  $10^{9.6}~M_{\sun}$, bottom $10^{10.6}~M_{\sun}$) and CC85 model velocities 800 \kms, 1800 \kms, and 2200 \kms left to right. $v = 0$ \kms $ = v_{escape}$ is the vertical blue line; gas to its left is bound by gravity, to right escapes.
The solid red curves trace the total mass per velocity bin. The dotted orange curve and the dot-dashed green curve show mass with temperatures in the Soft X-ray and Mid X-ray regimes respectively. The blue dashed curve shows cold mass at $<100$ K.\label{fig:massbins}}
\end{figure*}

\begin{figure*}[h!]
\centering
\includegraphics[height=.33\linewidth]{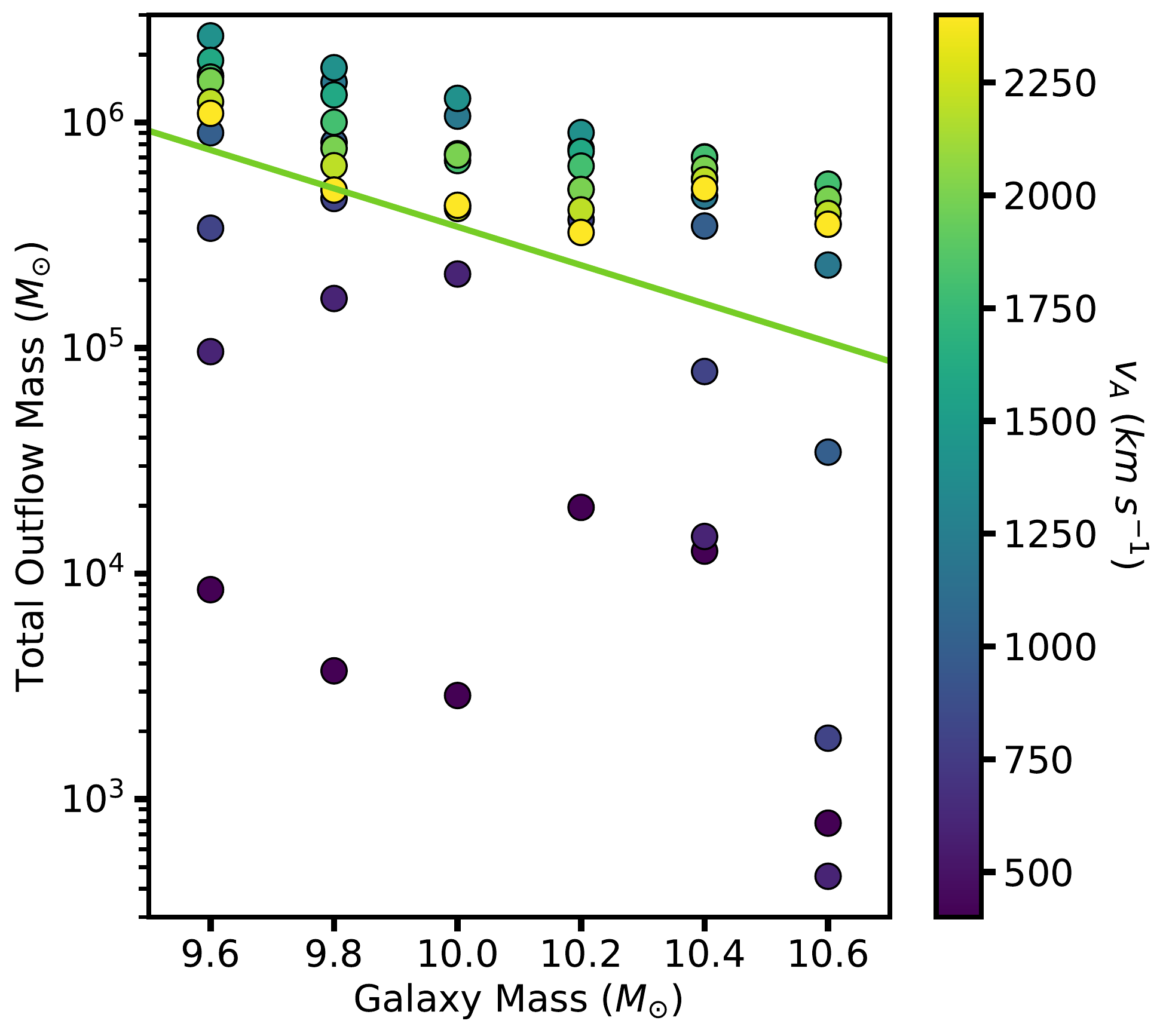}\quad\includegraphics[height=.33\linewidth]{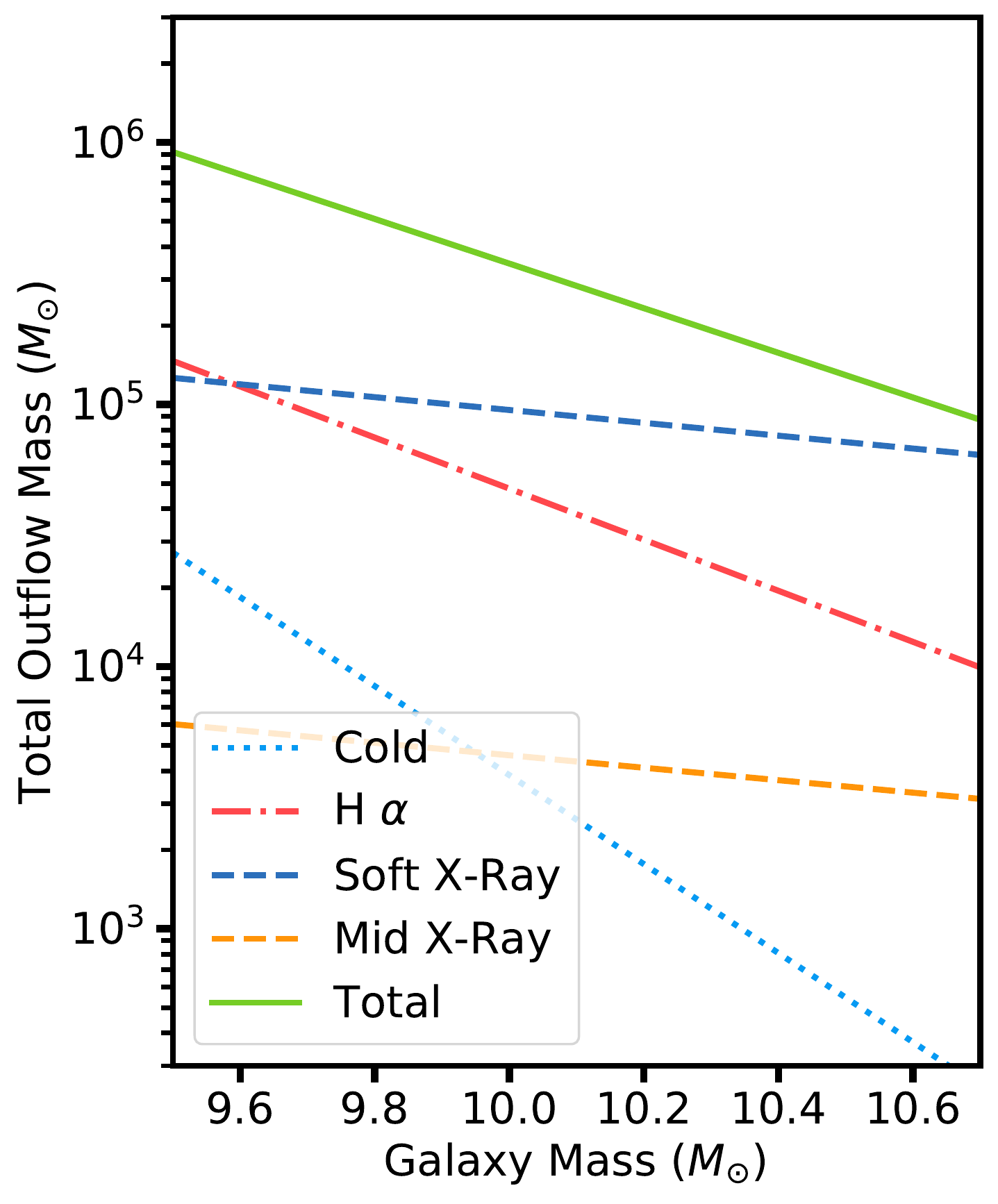}
\caption{Left plots the total mass above the escape velocity (everything to the right of the vertical lines in Figure \ref{fig:massbins}) with respect to galactic mass. All 66 simulations are plotted, and each dot is colored according to the CC85 model velocity. In solid green is the linear fit to the data with slope -0.85. At right I show the total mass in various gas temperature regimes from my simulations and their linear fits (fitted slopes given in Table \ref{tab:slopes}).\label{fig:outmass}}
\end{figure*}

Figure \ref{fig:massbins} shows mass distributions of six representative models. 
As expected, higher $v_A$ puts more mass at higher velocity, and a higher galaxy mass lowers total mass outflow for all temperature ranges. 
The majority of the outflow mass has temperatures in the Soft X-ray (0.5-3 keV) and Mid X-ray (3-10 keV) ranges.
This counters the expectation that warm gas dominates the outflow mass, but the hot gas only dominates in galaxies with mass $>10^{9.6}$ \msun.
Below this the warm gas dominates the outflow mass;
Figure \ref{fig:outmass} shows this cross over.

For each model I sum the total mass where $v > v_{escape}$ and plot it in Figure \ref{fig:outmass}. 
I fit the data using least squares minimization and find negative correlation between total outflow mass and galaxy mass with slope -0.85. 
Because constant mass is injected in my simulations, the total mass is directly proportional to the mass outflow rate. 
This allows direct comparison between the total outflow mass in my simulations and mass loading from a galaxy ($\eta$) as measured with the FIRE simulations of \citet{2015MNRAS.454.2691M}. 
Those authors found $\eta \propto v_{cir}^{-1}$ for galaxies $> 10^{9}$ \msun. 
From a set of galaxies with strong outflows \cite{2015ApJ...809..147H} obtained an inverse correlation slope -0.98. 
My measured slope is slightly less than either of these observations.
But when I consider only the mass of the warm gas (5,000-40,000 K) outflow, its slope -0.97 agrees with their and other results \citep{2013MNRAS.429.1922C,2015ApJ...809..147H,2015MNRAS.454.2691M,2019ApJ...886...74M,2019ApJ...886...29S}.

\begin{deluxetable}{lD}
\tablecaption{Slopes from linear fits to outflow mass vs.\ galaxy mass (Figure \ref{fig:outmass}.\label{tab:slopes})}
\tablehead{
\colhead{Temperature Regime} & \multicolumn2c{Slope}
}
\decimals
\startdata
Cold ($<100$ K) & -1.70 \\
H$\alpha$ (5,000 - 40,000 K) & -0.97 \\
Soft X-ray (0.5 - 3.0 keV) & -0.25 \\
Mid X-ray (3.0 - 10.0 keV) & -0.24 \\
Total Mass & -0.85 \\
\enddata
%\tablecomments{comment}
\end{deluxetable}

\begin{figure}[ht!]
\centering
\includegraphics[height=.4\linewidth]{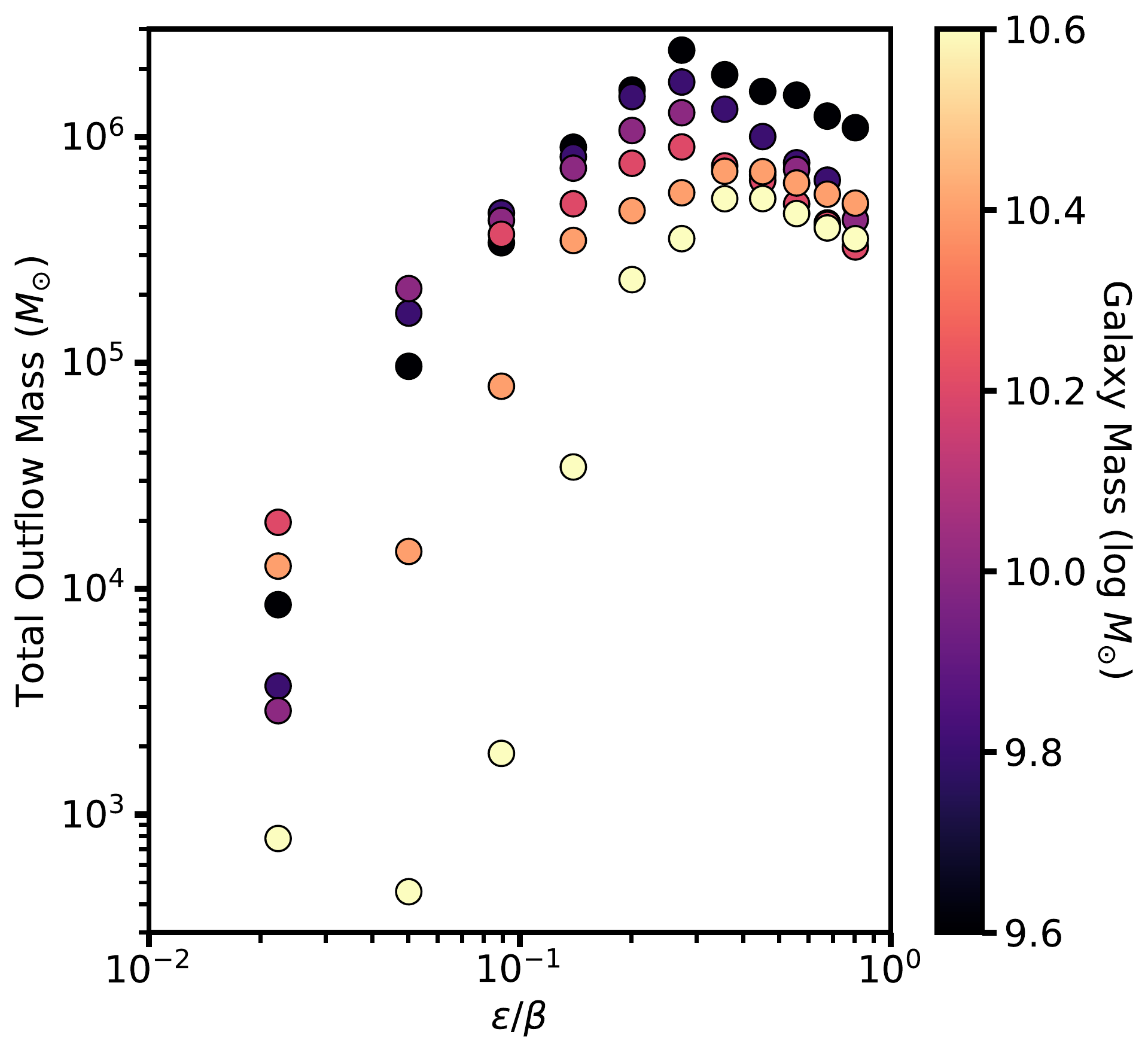}
\caption{The same total masses from Figure \ref{fig:outmass} plotted with respect to $\epsilon/\beta$ of each simulation using Equation \ref{eq:va}. The outflow velocity for the simulations increases from left to right. Circle coloring indicates the galactic mass for each simulation.\label{fig:sfemass}}
\end{figure}

Figure \ref{fig:outmass} right plots the best fits for the total outflow mass and the outflow mass in the cold, warm, and hot phases (split into Soft and Mid X-Ray temperature ranges). 
Both exhibit the same relation, but the Mid X-Ray range has $\sim 15\times$ lower mass. 
Table \ref{tab:slopes} reports slopes for all relations. Evidently, the mass outflow of hot gas does not follow the correlations of warm or cold gas. 
Thus, emission from warm gas cannot be a proxy for the outflow mass of the hot gas, especially in higher mass galaxies. 
Because the mass of Soft X-ray emitting gas dominates the outflow, estimates of the total mass outflow based on tracers of warm gas will significantly under-estimate the total mass of the outflow in higher mass galaxies, but is accurate in lower mass galaxies.

Figure \ref{fig:sfemass} plots the total mass of the outflow vs.\ the ratio of $\epsilon$ and $\beta$. 
This ratio measures the general efficiency of the starburst, and determines the velocity of the outflow using Equation \ref{eq:vg}. 
The relationship between $\epsilon/\beta$ and the total mass of the outflow is non-linear, peaking at $\epsilon/\beta= 0.2-0.4$. 
At these values the starburst is most efficient at removing gas from the galaxy. 
Assuming a fixed value for $\epsilon$ the ratio $\epsilon/\beta$ depends entirely on $\beta$. 
At lower $\beta$ the gas velocity may be higher, but less mass loading reduces the density and therefore the wind mass. 
At higher $\beta$ the wind may be denser, but moves slower so most gas will not escape the galaxy.

As noted in \S\ref{sec:setup:etabeta} for these simulations I fixed $\epsilon=1.0$, but the thermalization efficiency can have a range of values between 0.01 and 1.0 with the possibility of changing over time. 
It is possible to get the same outflow velocity with different values of $\epsilon$ and $\beta$ as long as the ratio is the same. 
This degeneracy prevents determining either $\epsilon$ or $\beta$ from just kinematic measurements. 
But multiwavelength observations of outflows can be used to determine the total galactic outflow rate ($\eta$) from IR, optical, UV, and X-ray luminosities. 
When combined with simulations can create a correlation between the measured luminosity of starbursts and the mass loading. 
This is currently and active area of research for the author.

\section{Maximum Galaxy Mass for Outflows}\label{sec:maxmass}

\begin{figure*}[ht!]
\centering
\includegraphics[height=.4\linewidth]{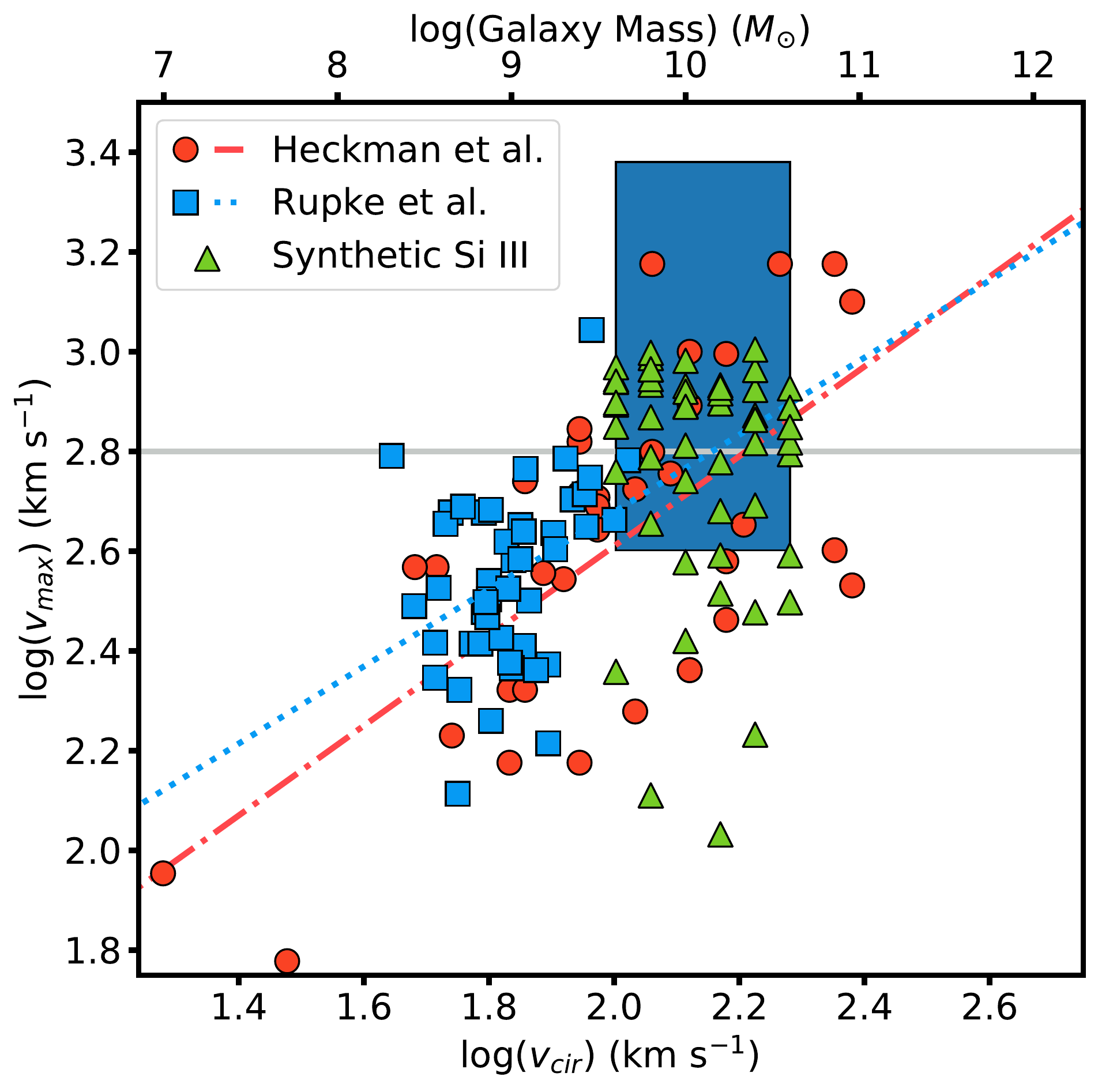}\includegraphics[height=.4\linewidth]{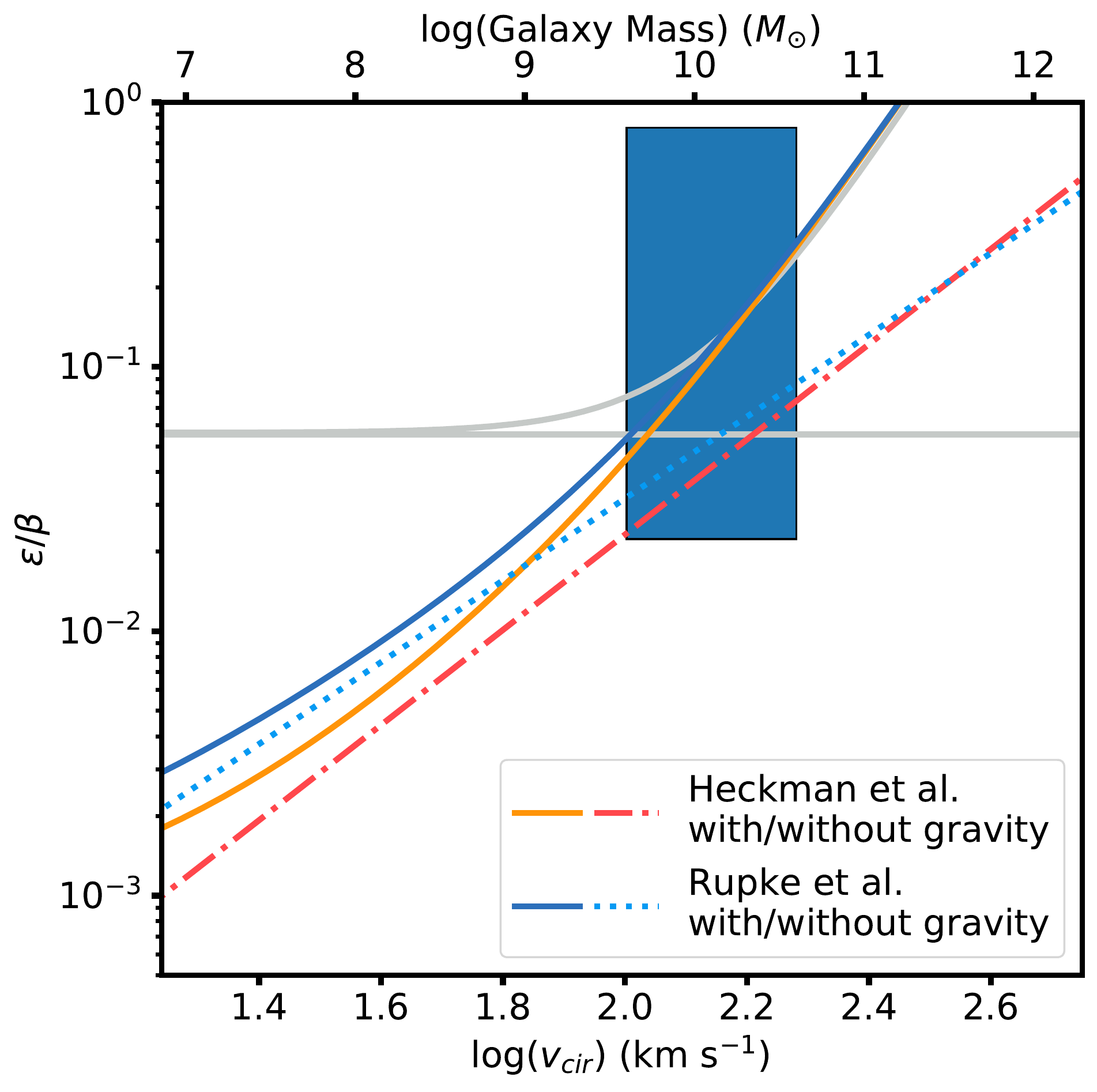}
\caption{The left panel plots maximum outflow velocity vs. circular velocity. Red circles are data from \cite{2016ApJ...822....9H}. Blue squares are data from \citet{2005ApJS..160..115R}. Lines show fits for both datasets. Green triangles show velocities from my simulations of the Si III lines. In both panels the blue rectangle shows the parameter space of my simulations. I have included a fit (grey line) at a constant $v_{max}$ for all $v_{cir}$. In the CC85 model this corresponds to constant $\epsilon/\beta$. At right I convert the best fits of Heckman et al.\ and Rupke et al.\ into $\epsilon/\beta$ values using the JA71 and CC85 models (with and without gravity, respectively). The two grey lines show $\epsilon/\beta$ for the constant $v_{max}$ as computed using the JA71 and CC85 models.\label{fig:epsbet}}
\end{figure*}

Observations show a correlation between outflow velocity and circular velocity of the host galaxy, for example
 \citet{2005ApJS..160..115R} and \cite{2016ApJ...822....9H} plotted in Figure \ref{fig:epsbet} together with the outflow velocity as measured by Si III lines from my models; both observations yield similar least-squares fit slopes 0.90 and 0.77, respectively.
Using the stellar mass set in each simulation and Tully-Fischer relation in \cite{2011MNRAS.417.2347R}, Figure \ref{fig:epsbet} also plots the circular velocity for each simulated galaxy using $v_{cir} = 0.278 \log M_{*} - 0.666$.

As noted in \S\ref{sec:setup:etabeta} my simulations may have nonphysical $\epsilon$ and $\beta$ values.
Hence I cannot establish their values individually, but can constrain their ratio from Equations \ref{eq:va} and \ref{eq:vg} by using the best fits from \cite{2016ApJ...822....9H} and \citet{2005ApJS..160..115R}. 
Physical values require (\citetalias{CC85}) $\epsilon/\beta < 1.0$; anything above would require thermalization efficiency $>1$ and/or a mass loading $<1$, representing significant mass-freeze out from stellar winds and supernova. 
This becomes my cut off point for starbursts that can form outflows. 
For both models, higher mass galaxies must have correspondingly higher $\epsilon/\beta$ to fulfill the observed $v_{max}$ vs.\ $v_{cir}$ relation.
From the fits of \cite{2016ApJ...822....9H} and \citet{2005ApJS..160..115R}, the simple model of \citetalias{CC85} allows for possible outflows from galaxies $\la 10^{13}$ \msun. 
That CC85 has no intrinsic limit for outflows to form in galaxies with mass $> 10^{12}$ \msun~ therefore requires
another mechanism to quench outflows such as AGNs or depletion of disk gas.

Alternatively, the JA71 model cuts off just above $10^{11}$ \msun; beyond, outflows require nonphysical values of $\epsilon$ and $\beta$. 
This allows starbursts to quench their own outflows in galaxies with mass $10^{11-12}$ \msun~without invoking an AGN or gas depletion. 
Thus, the JA71 model yields $\approx 10^{11.5}$ \msun~as the maximum allowable galaxy mass for outflows. 

In Figure \ref{fig:epsbet} the grey horizontal lines at constant $v_{max}$ for all $v_{cir}$ is
constant $\epsilon/\beta$ from the CC85 model. 
The line for JA71 is identical for galaxies $<10^{9}$~\msun~but quickly diverges from CC85 for higher masses to produce the same maximum mass as Heckman et al.\ and Rupke et al.\ for $\epsilon/\beta = 1.0$. 
The mass limit of Equation \ref{eq:vg}, being robust to a broad range of $v_{max}$ vs. $v_{cir}$ slope fits, is not an artifact of galaxy selection.

\section{Discussion}\label{sec:discussion}

%Here I discuss the implications of my results for combining analytic models with 3D simulations and the need for multiwavelength observations of starburst driven galactic outflows.

\subsection{Dynamics of a Three Phase Outflow}\label{sec:discussion:one}

\citetalias{Tanner17} examined how the velocity of the three-phase outflow changes by the SFR. 
For a single galaxy mass $6\times 10^9$~\msun~the CC85 model with radiative cooling sufficed to predict the velocity of the hot gas, but could only predict the velocities of warm and cold phases at a high SFR. 
In the present work with the SFR constant, I tested how the outflow velocity depends on the stellar mass of the galaxy, finding
that the CC85 model is inadequate to describe the outflow velocity of hot gas for stellar mass $> 10^{10}$ \msun. 
The JA71 model -- the CC85 model plus a gravitational potential -- is needed to accurately model the velocity of the hot gas. This agrees with the analytic models of \citet{2020MNRAS.492.3179Y}, and disagrees with \citet{2020arXiv200210468S} who assumed that gravity does not affect that phase. 

All recent hydrodynamical simulations include gravity, but analytic models that analyze model results often neglect it or assume it to be insignificant because an M82-mass galaxy is typically used \citep{CooperI,2009ApJ...698..693F,Creasey,2013MNRAS.430.3235M,2013MNRAS.434.3572R,2014MNRAS.441..431S,2016ApJ...821....7T,Tanner17,2020MNRAS.492.3179Y,2020arXiv200210468S} where the difference between the CC85 and the JA71 models is small compared to other effects.
However, my results show that for simulations with stellar mass $> 10^{10}$ \msun~gravity can dominate the dynamics of the hot gas, especially for higher density (high $\beta$) winds. 
Evidently, hydrodynamical simulations of outflows should not just assume an M82-mass galaxy, but also examine more massive galaxies. 

For sufficiently high values of $\beta$ the JA71 model also applies to the warm gas, but for $\beta < 3$ (corresponding to a hot wind speed of $>1400$ \kms and a warm wind speed of $> 900$ \kms) it does not appear that the JA71 model adequately describes the dynamics of the warm phase. 
Recent approaches model neutral or slightly ionized gas as clouds driven by ram pressure \citep{2015ApJ...809..147H,2019ApJ...878...84M} rather than a gas shell driven by the starburst as CC85 and JA71 assumed.
The cloud+ram pressure model has successfully explained observations of cold, dense clouds in a hot wind \citep{2009ApJ...703..330C,2016ApJ...821....7T} that ablate mass to contribute to the warm phase. 
However, simulations have also shown that ram pressure will disrupt clouds so cannot account for the mass of warm gas observed in the winds \citep{2016ApJ...822...31B,2017ApJ...834..144S,2017MNRAS.468.4801Z,2020arXiv200210468S}. 
This discrepancy highlights the need for simulations having different wind geometry and refined physics. 

Finally, my results show in Figures \ref{fig:linev} and \ref{fig:outvel} that the gas outflow has three distinct phases with different velocities. 
Factors that set each phase include mass loading ($\beta$), gravity, SFR, and wind geometry. 
For example, $\beta$ sets the outflow velocities of all three phases, the SFR only affects the warm and cold phases, whereas gravity has biggest effect on the hot phase. 
Each phase should be considered separately with its own analytic model as done recently \citep{2015ApJ...809..147H,2019ApJ...878...84M,2020arXiv200210468S}.

The values of $\beta$ that I have worked with here should be considered as upper limits. 
Because I set $\epsilon=1.0$, more realistic values of $\epsilon$ would require lower values of $\beta$ to produce the same hot gas velocity.
As $\beta$ has a lower limit of 1.0, the value of $\epsilon$ will determine the upper limit for the hot gas velocity from starburst driven winds.

\subsection{Outflow Mass}\label{sec:discussion:two}

When considering simulations with the same mass loading, more massive galaxies have less total mass outflow than less massive galaxies because of their deeper gravitational potential. 
The correlation of total outflow mass vs. galaxy mass that I find has shallower slope than that observed and in other simulations \citep{2013MNRAS.429.1922C,2015ApJ...809..147H,2015MNRAS.454.2691M}. 
But summing only the gas mass at peak H$\alpha$ emission (5,000-40,000 K) matched the observed slope $\approx-1$. 

This correlation is used in subgrid models of cosmological simulations \citep{2003MNRAS.339..312S,2013ApJ...770...25A,2015MNRAS.454.2691M,2018MNRAS.473.4077P,2020MNRAS.493....1H}. 
But as Figure \ref{fig:outmass} shows, its slope only holds for the warm phase. 
The slope of the hot phase is much shallower, $\approx -0.25$. 
Thus not only are their velocities potentially quite different between warm and hot phases (\S\ref{sec:velocity}), but galactic mass loading and gas mass differ for the warm and hot phases. 
For galaxies with $>10^{10}$ \msun~the hot gas dominates the outflow mass, becoming by
$\sim 10^{11}$~\msun~ $\sim10\times$ more massive than the warm phase. 
Thus the hot and warm gas outflows should be considered separately in the subgrid models of cosmological simulations. 
Moreover, negative correlation $-1$ was established using H$\alpha$ tracers and similar optical/UV emission. 
The correlation for the hot X-ray emitting gas should be confirmed through similar X-Ray measurements. 

In this paper I focus solely on the slope or relationship between the total outflow mass and stellar mass. 
The actual amount of mass flowing out of the galaxy would depend on the SFR, $\epsilon$, $\beta$, and the ambient pressure.
As shown in Figure \ref{fig:sfemass} the outflow mass is maximized for a certain range of $\epsilon/\beta$. 
Because multiple values of $\epsilon$ and $\beta$ can produce the same value for $\epsilon/\beta$, further work will be needed to disentangle this degeneracy.
But a lower value of $\epsilon$ will require a corresponding lower value of $\beta$ to achieve the same wind velocity. 
This will decrease the total mass in the outflow.

While I have fixed the SFR at 50 \msyr, a higher or lower SFR will raise or lower the total mass in the wind. 
The exact relation between the mass outflow and the SFR would have to be considered for each of the three gas phases separately. 
The hot gas mass should have a direct relation to the SFR, but the outflow mass in warm and cold phases will have a much more complex relation but should also increase with increasing SFR.
This would also need further study.

\subsection{Limiting Efficiencies}\label{sec:discussion:three}

As explained in \S\ref{sec:setup:etabeta} some values of $\epsilon$ and $\beta$ that I chose can be nonphysical. 
Observations \citep{2005ApJS..160..115R,2016ApJ...822....9H} show that outflow velocity increases with galaxy mass. 
In contrast, results in \S\ref{sec:velocity} show that the outflow velocity should decrease with increasing galaxy mass. 
This decrease only occurs at constant $\epsilon$ and $\beta$, but their values can evolve over time so $\epsilon/\beta$ can vary with galaxy mass. 
Given the fits from \cite{2016ApJ...822....9H} and \citet{2005ApJS..160..115R} in Figure \ref{fig:epsbet}, when calculating $\epsilon/\beta$ using Equations \ref{eq:va} and \ref{eq:vg}, $\epsilon/\beta$ clearly increases with galaxy mass. 
Thus thermalization efficiency would be higher and/or the mass loading would be lower in higher mass galaxies. 
This represents a fundamental shift in thinking about starburst efficiencies in galaxies of different mass. 

For $\epsilon/\beta$ to increase either $\epsilon$ must increase or $\beta$ must decrease, or both. 
For the highest mass galaxies the values of $\epsilon$ and $\beta$ must converge to 1.0 to maintain an outflow. 
This would require that all radiation from the starburst be absorbed by the ISM, and that the SFE be as high as possible. 
Physical limitations of the thermalization and star formation efficiencies will set an upper stellar mass limit at which galaxies can form outflows.

The CC85 model shows that $\epsilon/\beta$ must increase linearly to increase velocity as galactic mass increases.
However, the divergence of the JA71 model from the CC85 model at $>10^{10}$ \msun~ shows that the fits from \cite{2016ApJ...822....9H} and \citet{2005ApJS..160..115R} would require unrealistic values of the thermalization efficiency and mass loading for galaxies $>10^{11.5}$ \msun. 
Because the data scatter, it would be possible to have outflows with a galaxy mass $>10^{11.5}$ \msun, but by $10^{12}$ \msun~all outflows should be quenched from the limit on starburst efficiency.

This result explains the scarcity of outflows in galaxies of $>10^{11}$ \msun, where outflow would require no loss of radiative energy and minimal mass loading into the wind beyond that from supernova ejecta and stellar winds. 
How adding cosmic rays, radiation pressure, magnetic fields, and other effects would influence this mass limit must be investigated.
But application of the JA71 model shows that starburst driven galactic outflows will quench somewhere between $10^{11}$ and $10^{12}$ \msun~even without an AGN. 
This only explains quenching of outflows and not of star formation, although it is possible that the mechanisms that increase $\epsilon/\beta$ in high mass galaxies also quench star formation. 
While AGN certainly have a critical role in regulating outflows, the JA71 model suggests that starburst-driven outflows can self regulate. 

\section{Conclusions}\label{sec:conclusions}

\begin{enumerate}
    \item The model of \citetalias{CC85} for outflows is accurate for low mass galaxies, but the gravity at higher masses greatly affects outflow kinematics. Analytic models of outflows with $>10^{10}$ \msun~must include gravity. The  \citetalias{JA71} model should be used for galaxies more massive than M82 \citep[or equivalent][]{2020MNRAS.492.3179Y} to estimate the kinematics of hot gas but not necessarily those of outflowing warm and cold gas.
    \item Outflow kinematics differ for each of the three phases, so should be considered separately in analytic models. 
    \item Using M82 as the canonical model in many 3D hydrodynamical simulations undervalues the influence of gravity when comparing simulations to analytic models and observations. Either a higher mass galaxy like NGC 3079 should be used, or a mix of high and low mass galaxies.
    \item The mass outflow, or galactic mass loading, vs.\ galaxy mass relation differs for different gas phases. The hot gas has a relatively shallow relation, the warm gas steeper, and cold gas even steeper. Thus outflows from high mass galaxies are almost entirely of hot gas.
    \item Different dynamics of the three phases emphasizes the need for multi-wavelength observations. Properties of neutral or slightly ionized gas are not necessarily relevant to those of highly ionized gas. 
    \item When interpreting observational fits, the \citetalias{JA71} model yields physically realistic values of thermalization efficiency and mass loading of outflows up to galaxy mass $\sim10^{11.5}$ \msun, independent of the SFR. Above this, outflows are prohibited by excessive efficiencies required of the starburst.
\end{enumerate}

%Hopefully the recommendation to reconsider using only M82 as the canonical galaxy for 3D hydro simulations will result in a broader range of galaxy masses considered in simulations. 

\acknowledgments
I acknowledge the support of Gerald Cecil and his comments that improved this paper. 
My code and analysis methods were developed with support through our NASA Herschel grant NHSC-OT-1-1436036 (P.I. S. Veilleux).
I also acknowledge Tom Crute at Augusta University and Kim Weaver of GSFC for support and encouragement. 

\appendix

\section{Outflow Velocities}\label{app:outvel}

\begin{figure}
    \plotone{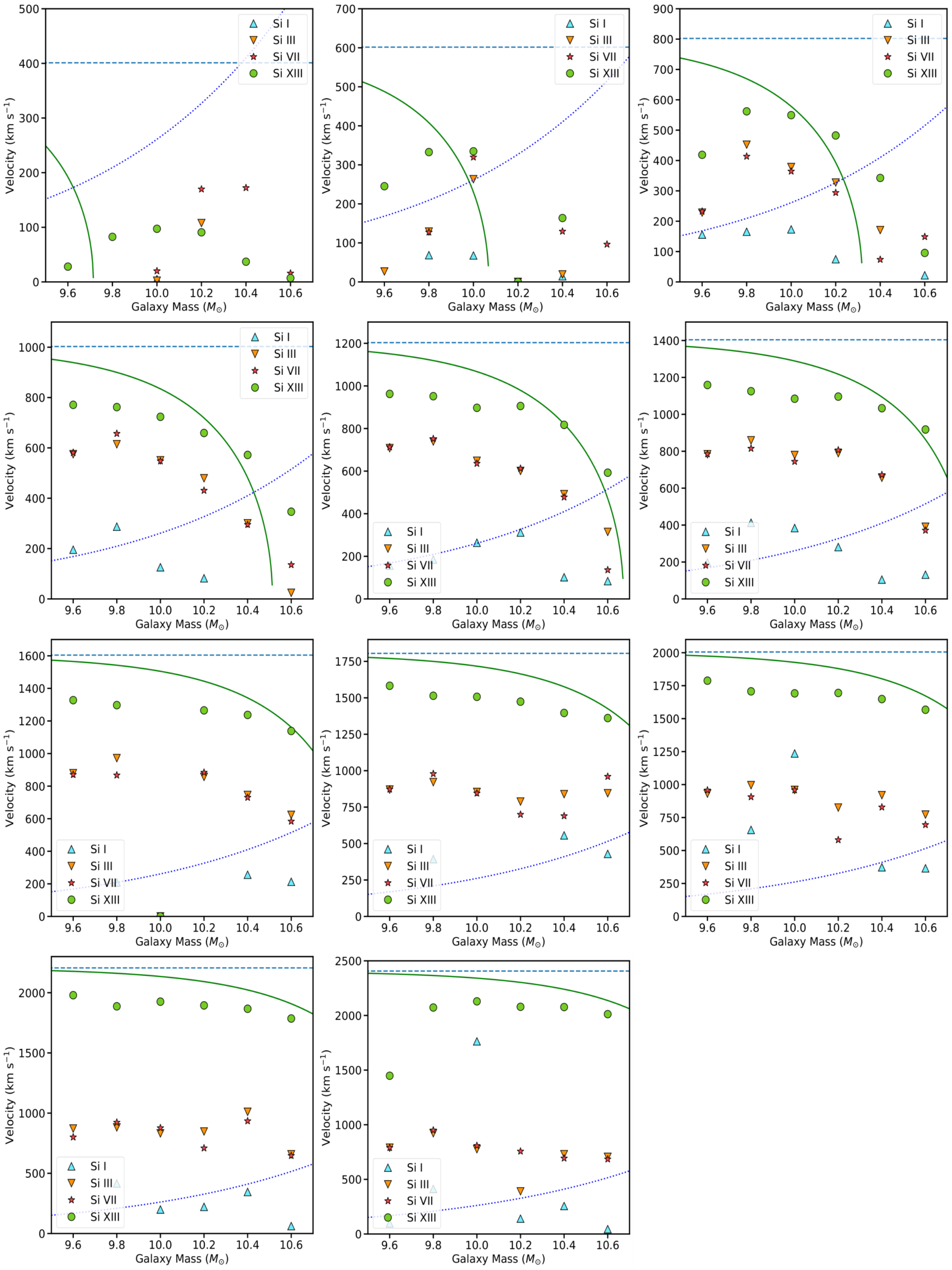}
    \caption{As for Figure \ref{fig:linev} I plot outflow velocity vs. galaxy mass. Outflow velocities are measured by Si I (upward triangles), Si III (downward triangles), Si VII (stars), and Si XIII (circles). In each panel the blue dashed line indicates the CC85 model velocity ranging from 400 \kms (top left) to 2400 \kms (bottom center). The solid green curve is the CC85 model velocity modified to include the gravitational potential, and the dotted purple curve is the escape velocity for gas in the galactic nucleus.\label{fig:outvel}}
\end{figure}

\bibliography{paper3}{}
\bibliographystyle{aasjournal}

\end{document}